\documentclass[twocolumn,showpacs,preprintnumbers,amsmath,amssymb]{revtex4}

\usepackage{amsmath,amssymb}
\usepackage{mathrsfs}
\usepackage{ulem}
\usepackage[dvips]{graphics,color}
\usepackage{dcolumn}
\usepackage{bm}

\newenvironment{ul}[1]{
\begin{itemize}
\setlength{\partopsep}{#1pt}
\setlength{\topsep}{#1pt}
\setlength{\itemsep}{#1pt}
\setlength{\parsep}{#1pt}
\setlength{\parskip}{#1pt}}{
\end{itemize}
}

\newcommand{\bvec}[1]{\mbox{\boldmath $#1$}}

\begin{document}


\title{Chaotic itinerancy and power-law residence time distribution \\
in stochastic dynamical system}
\author{Jun Namikawa}
\email{jnamika@jaist.ac.jp}
\affiliation{
School of Knowledge Science, \\
Japan Advanced Institute of Science and Technology, \\
Tatsunokuchi, Ishikawa, 923-1292, Japan
}%
\date{\today}%

\begin{abstract}

To study a chaotic itinerant motion among varieties of ordered states, we propose a stochastic model based on the mechanism of chaotic itinerancy.
The model consists of a random walk on a half-line, and a Markov chain with a transition probability matrix.
To investigate the stability of attractor ruins in the model,  we analyze the residence time distribution of orbits at attractor ruins.
We show that the residence time distribution averaged by all attractor ruins is given by the superposition of (truncated) power-law distributions, if a basin of attraction for each attractor ruin has zero measure.
To make sure of this result, we carry out a computer simulation for models showing chaotic itinerancy.
We also discuss the fact that chaotic itinerancy does not occur in coupled Milnor attractor systems if the transition probability among attractor ruins can be represented as a Markov chain.

\end{abstract}


\maketitle

The concept of chaotic itinerancy (CI) was proposed as a universal class of dynamics with multiattractor, such as a chaotic itinerant motion among varieties of ordered states\cite{IOM1989,K1990,T1991}.
CI is generally described as follows:
Dynamical orbits are attracted to an ordered motion state, and stay there for a while.
Subsequently they separate from the ordered state, and enter into high-dimensional chaotic motion.
By those repetitions, they successively itinerate over ordered motion.
Each of such ordered motion states is called an ``attractor ruin''.
Showing mathematical properties of CI has become an important problem in a variety of systems from many scientific disciplines, including semiconductor physics, chemistry, neuroscience, and laser physics\cite{KT2003}.

To mathematically characterize CI, some researchers suggest that attractor ruins can be represented as Milnor attractors\cite{K1997,K1998,TU2003,KT2003}.
A Milnor attractor is defined as a minimal invariant set that has a positive measure as its basin of attraction\cite{M1985}.
Since this definition does not exclude the possibility that the orbits leave from any neighborhood of the attractor, attractor ruins may be described with Milnor attractors.
Indeed, in several models the existence of Milnor attractors with CI was reported\cite{K1997,K1998,TU2003}.
However, it is still unclear whether Milnor attractors exist in a system whenever the system shows CI.

The stability of attractor ruins is one of the important properties by which CI is characterized.
Residence time distribution is regarded as the statistical property of such stability.
In this paper we investigate the residence time distribution of orbits at attractor ruins, and discuss the possibility of CI that occur in a dynamical system containing Milnor attractors.
To describe the distribution, we argue a mechanism of transition among attractor ruins, and we propose a model based on that mechanism.

As an example of a dynamical system showing CI, we first review the globally coupled map (GCM)\cite{K1990} given by
\begin{equation} \label{equation:globally_coupled_map}
x_{t+1}(i) = (1 - \epsilon)f(x_{t}(i)) + \frac{\epsilon}{N}\sum_{j=1}^{N}f(x_{t}(j)),
\end{equation}
where $x_t(i)$ is a real number, $t$ is a discrete time step, $i$ is the index for elements ($i = 1,2,\cdots,N = \textrm{system size}$) and $f$ is a map on $\mathbb{R}$ as the local element in Eq.(\ref{equation:globally_coupled_map}).
When the elements $i$ and $j$ are synchronized, i.e., $x_t(i)\approx x_t(j)$, they are in a cluster.
Each element belongs to a cluster, including a cluster with one element.
The state of GCM is characterized by a partition of elements into clusters.
The state of partition is called clustering condition.
CI of GCM is observed as chaotic changes of the clustering conditions\cite{K1997,K1991}.
Namely, in GCM a clustering condition is the connotation of an attractor ruin.
If elements of each cluster in a clustering condition completely synchronize, that is, $x_t(i) = x_t(j)$ for any elements in one cluster, the clustering condition is an invariant subspace.
Thus, CI of GCM is considered as a phenomenon that orbits approach an invariant subspace after staying in the neighborhood of another invariant subspace for some time.
As an indicator of the stability of clustering conditions, the local splitting exponent $\lambda_{\mathrm{spl}}^T(i,n)$\cite{K1994} is defined by 
\begin{equation} \label{equation:local_splitting_exponent}
\lambda_{\mathrm{spl}}^T(i,n) = \frac{1}{T}\sum_{m=n}^{n+T}\log|(1-\epsilon)Df(x_{m}(i))|.
\end{equation}
The local splitting exponent $\lambda_{\mathrm{spl}}^T(i,n)$ represents a local expansion rate of a separation distance between the element $i$ and an adjacent element of $i$.
Thereby we can consider that $\lambda_{\mathrm{spl}}^T(i,n)$ represents a local expansion rate of a distance between two elements in the cluster to which the element $i$ belongs.
Indeed, the element $i$ is contained in a cluster with more than one element if $\lambda_{\mathrm{spl}}^T(i,n)$ is negative, and the element does not synchronize with any other elements if $\lambda_{\mathrm{spl}}^T(i,n)$ is positive.
Hence if the sign of $\lambda_{\mathrm{spl}}^T(i,n)$ switches, then a clustering condition changes.
Moreover, if an orbit sufficiently approaches an invariant subspace corresponding to a clustering condition, the local splitting exponent can describe the distance between the orbit and the invariant subspace by using a logarithmic scale.

In general, if the system is unstable in a direction normal to an invariant subspace, an orbit leaves an attractor ruin containing the invariant subspace.
Note that the mechanism which destabilizes an invariant subspace is called a blowout bifurcation, and is known to cause bursts in on-off intermittency\cite{OS1994}.

In summary, the mechanism of CI can be described as follows:
\begin{ul}{1}
\item An orbit leaves an attractor ruin, when the distance between the orbit and an invariant subspace in the attractor ruin is greater than a certain value.
\item If nonlinearity, i.e., a local expansion rate, is weak in a direction normal to the invariant subspace, the distance between the orbit and the invariant subspace decreases.
Otherwise the distance increases.
\end{ul}

Based on the mechanism mentioned above, we introduce a prototype model in which CI occurs.
We consider the distance between a orbit and the nearest invariant subspace.
While in GCM the number of values for nonlinearity, i.e., local splitting exponents, is equal to the system size, we consider only one variable to represent nonlinearity.
Besides, the variable is a stochastic value of $1$ or $-1$, and is decided by a probability associated with the nearest attractor ruin to the orbit.
Furthermore, we introduce a probability to govern transition among attractor ruins.
In definition of the transition probability, we assume that the influence of a past attractor ruin upon the transition decays rapidly, such that the transition depends on only finite past attractor ruins.
This assumption provides the simplest case as the chaotic itinerant dynamics on attractor ruins in CI.

\begin{figure*}[hbtp]
\scalebox{0.85}{\includegraphics{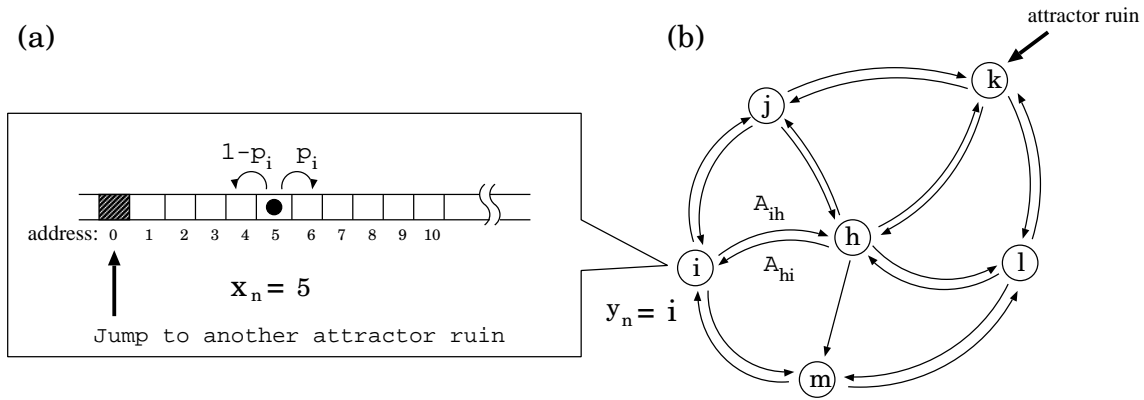}}
\caption{\small Illustration of an itinerant dynamics model.
(a)Schematic view of the space of $x_n$.
The stability of an attractor ruin is represented by a random walk on a half-line, where $p_i$ is a probability that a point moves to right on the half-line, i.e., $x_{n+1} = x_n + 1$.
If the point arrives on the leftmost address on the half-line, transition among attractor ruins happens.
(b)Schematic view of the space of $y_n$.
Transition among attractor ruins is described by a Markov chain with a transition probability matrix $A$, where $A_{i,j}$ is a probability moving from attractor ruin $i$ to $j$, i.e., $y_{n} = i$ and $y_{n+1} = j$.
}
\label{figure:ci_like_model}
\end{figure*}

We define a stochastic model satisfying the characteristics in the previous paragraph.
Let $n \in \mathbb{N}$, $x_n \in \mathbb{N} \cup \{0\}$, and $y_n = \{1,\cdots,M\}$.
A series of positive integers $\{x_n\}_{n=0}^{\infty}$ is defined by
\begin{equation}
x_{n+1} = \left\{ \begin{array}{cc}
0 & \mbox{$x_n = 0$ and $\epsilon_n = -1$}, \\
x_n + \epsilon_n & \mbox{otherwise},  \\
\end{array} \right.
\end{equation}
where $\epsilon_n$ is a stochastic variable either $1$ with probability $p_{y_n}$ or $-1$ with $1-p_{y_n}$.
In other words, $x_n$ is given by a random walk on a half-line(see Fig.\ref{figure:ci_like_model}(a)).
A series $\{y_n\}_{n=0}^{\infty}$ is also defined by
\begin{equation}
y_{n+1} = \left\{ \begin{array}{cc}
y_n & \mbox{$x_n > 0$},  \\
z_n & \mbox{otherwise}, \\
\end{array} \right.
\end{equation}
where $z_n$ is a stochastic variable such as $z_n = k$, with probability $A_{{y_{n}}k}$ being the element of a non-negative square matrix $A$.
Namely, $y_n$ is determined by a Markov chain with the transition probability matrix $A$(see Fig.\ref{figure:ci_like_model}(b)).
Here $y_n$ is the index for attractor ruins, and $e^{-x_n}$ is the distance from an invariant set corresponding to $y_n$.
In addition, $p_{y}$ denotes the intensity of nonlinearity on attractor ruin $y$, and $A_{ij}$ is the transition probability from attractor ruin $i$ to $j$.

\begin{figure}[hbtp]
\scalebox{0.6}{\includegraphics{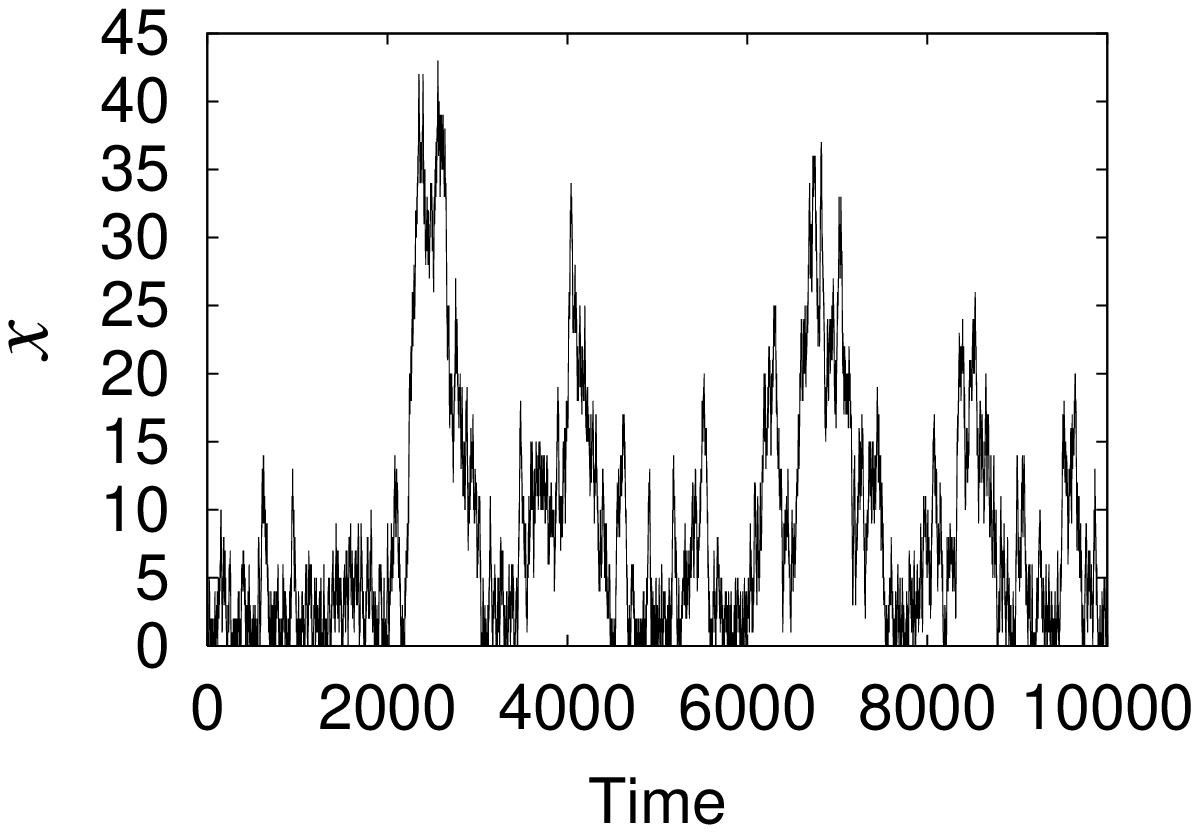}}

\scalebox{0.6}{\includegraphics{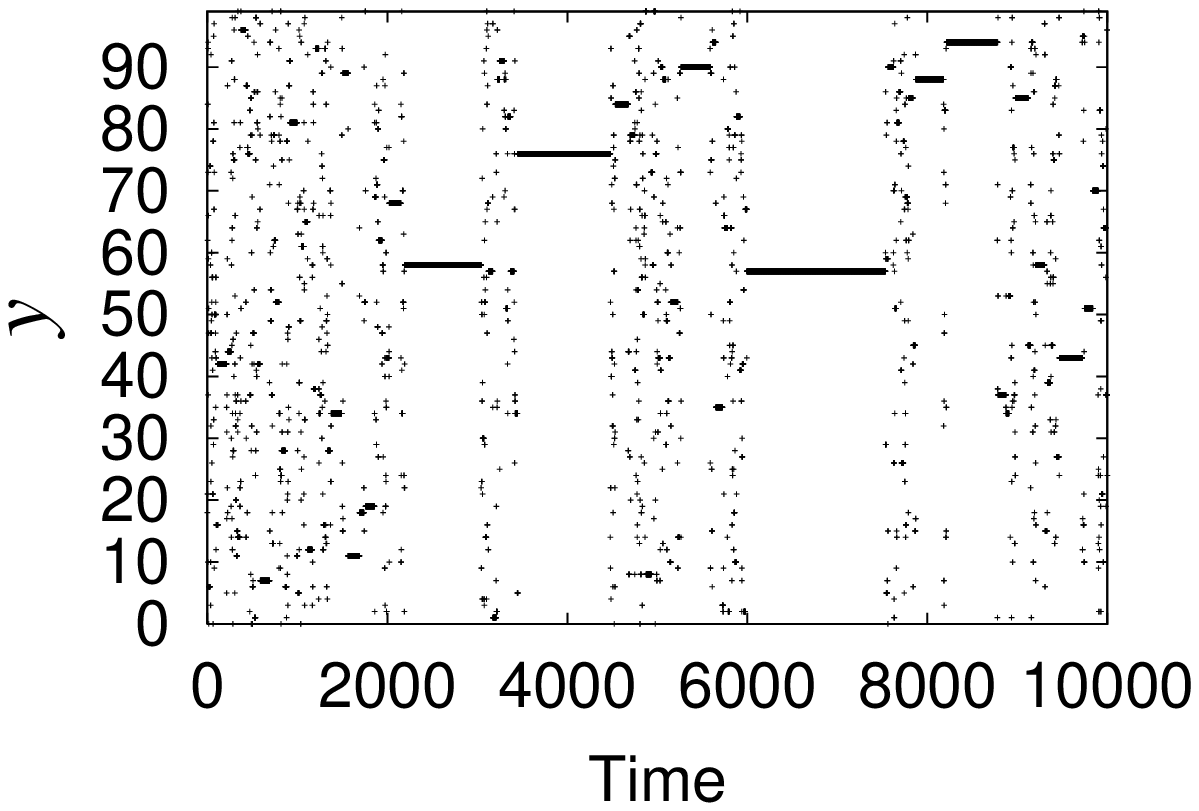}}
\caption{\small Time series of $x_n$ and $y_n$. Here, $M = 100$, $A_{ij} = \frac{1}{M}$, and $p_i = 0.45 + \mu_i$ where $\mu_i$ is a random number taken from $[-0.05,0.05]$.}
\label{figure:example_of_dynamics}
\end{figure}

In Fig.\ref{figure:example_of_dynamics}, we display time series of $x_n$ and $y_n$ with respect to $M = 100$, $A_{ij} = \frac{1}{M}$, and $p_i = 0.45 + \mu_i$ where $\mu_i$ is a random number taken from $[-0.05,0.05]$.
In Fig.\ref{figure:example_of_dynamics}, we can see both the region where $y_n$ is fixed, and the region where $y_n$ is dynamically changed.
Moreover, in the region where $y_n$ is fixed, $x_n$ is large, i.e., the orbit approaches the invariant set.

\begin{figure}[hbtp]
\scalebox{0.6}{\includegraphics{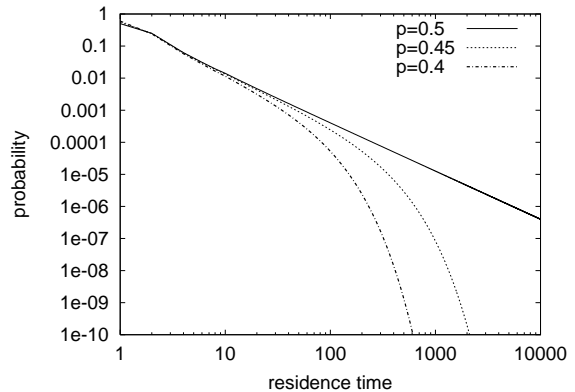}}
\caption{\small The residence time distribution at the attractor ruin $i$ with probability $p_i$.}
\label{figure:local_residence_time_distribution}
\end{figure}

Let us introduce the residence time distribution of orbits at an attractor ruin in our model.
The probability $P(i,t)$ for residence time $t$ at an attractor ruin $i$ is given by the probability for $x_{n+t} = 0$ if $x_n = 0$,  $y_n = i$, and $x_{n+k} > 0$ for any $k < t$ and $n \in \mathbb{N}$.
Since $t$ is regarded as the recurrent time to origin in one-dimensional random walk\cite{F1957,S1976}, this probability is given by
\begin{equation} \label{equation:recurrent_time_to_origin}
P(i,t) = \left\{ \begin{array}{cc}
1-p_i & t = 1, \\
\frac{[p_i(1-p_i)]^{n}}{n}{{2n-2}\choose{n-1}} & \exists n \in \mathbb{N} ~ t = 2n, \\
0 & \mbox{otherwise}. \\
\end{array} \right.
\end{equation}
If $t = 2n$ and $n \in \mathbb{N}$, $\log P(i,2n)$ can be denoted as
\begin{eqnarray} \label{equation:power_law}
\log P(i,2n) & \sim & n\log[p_i(1-p_i)] - \frac{1}{2}\log\big(\pi n^2 (n-1)\big) \nonumber \\
 & & + (2n-2)\log 2 
\end{eqnarray}
given by the approximate expression of Eq.(\ref{equation:recurrent_time_to_origin}).
If $p_i = 0.5$, then the right hand side of Eq.(\ref{equation:power_law}) takes the form of a constant plus the second term.
Namely, the residence time at the attractor ruin $i$ is governed by power-law distribution.
If $p_i < 0.5$, then $n$ in the first and third terms in Eq.(\ref{equation:power_law}) are not canceled, so that residence time distribution is truncated (see Fig.\ref{figure:local_residence_time_distribution}).
The probability that the residence time at the attractor ruin $i$ is longer than $t$ is denoted by
\begin{equation}
Q(i,t) = 1 - \sum_{k=1}^{t}P(i,k).
\end{equation}
It is easy to see that $Q(i,1) = p_i$ and $Q(i,2n+1) = Q(i,2n)$ for any $n \in \mathbb{N}$.
Since 
\begin{equation}
P(i,2n) = \frac{[p_i(1-p_i)]^{n}}{2} \Big[ 4{{2n-2}\choose{n-1}}-{{2n}\choose{n}} \Big],
\end{equation}
then 
\begin{eqnarray}
Q(i,2n) & = & [1-4p_i(1-p_i)]\sum_{k=1}^{n-1}\frac{[p_i(1-p_i)]^k}{2}{{2k}\choose{k}}  \nonumber \\
 & & + p_i-2p_i(1-p_i) + \frac{[p_i(1-p_i)]^{n}}{2}{{2n}\choose{n}}.
\end{eqnarray}
Hence $\lim_{t\rightarrow \infty}Q(i,t) = 0$ if $p_i \leq 0.5$, and $\lim_{t\rightarrow \infty}Q(i,t) > 0$ if $p_i > 0.5$.
Accordingly, while transition from the attractor ruin $i$ surely takes place when $p_i \leq 0.5$, transition does not necessarily take place when $p_i > 0.5$.

Next, we investigate the residence time distribution averaged by all attractor ruins.
In studying CI, the residence time distribution averaged by all attractor ruins is more useful than that of each attractor ruin.
This is because the distribution of an individual attractor ruin can be studied only when the structure of the invariant sets is clear, but showing the structure is difficult in high-dimensional dynamical systems.

To simplify the discussion here, we assume without loss of generality that transition probability matrix $A$ is irreducible\footnote{A non-negative square matrix $A$ is irreducible iff for any $1 \leq i, j \leq M$ there exists $n \in \mathbb{N}$ such as $A^n_{ij} > 0$, where $A^n$ is the $n$th power of $A$.}.
When the eigenvector associated with a positive maximum eigenvalue (which exists by the Perron-Frobenius theorem) is denoted as $\bvec{r}$, normalization of $\bvec{r}$ denoted by $\bvec{q}$ represents stationary distribution in a Markov chain with the transition probability matrix $A$.

\begin{figure}[hbtp]
\scalebox{0.6}{\includegraphics{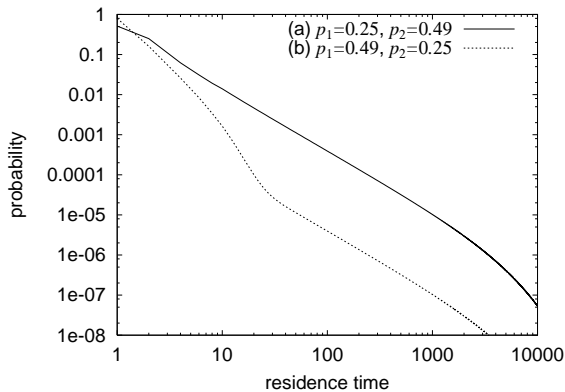}}
\caption{\small The residence time distribution averaged by all attractor ruins for $M = 2$, $q_1 = 0.01$, and $q_2 = 0.99$.
The case of $p_1 = 0.25$ and $p_2 = 0.49$ is shown by the solid line (a), and the case of $p_1 = 0.49$ and $p_2 = 0.25$ is drawn by the broken line (b).
}
\label{figure:global_residence_time_distribution}
\end{figure}

Suppose that $p_i \leq 0.5$ for any $1 \leq i \leq M$.
Since transition from each attractor ruin always happens when the probability is one, orbits itinerate over all attractor ruins.
Hence the probability of residence time $t$ for any attractor ruin is described by $\sum _{i=1}^{M} q_i P(i,t)$ from probability $P(i,t)$ and stationary distribution $\bvec{q} = \{q_i\}_{i=1}^{M}$.
Consequently, the residence time distribution averaged by all attractor ruins is the superposition of (truncated) power-law distributions.
As a result, there is a case in which such distribution does not seem to follow power-law.
In Fig.\ref{figure:global_residence_time_distribution}, we show examples of the residence time distribution with two attractor ruins.
In the case of line (a), the residence time distribution follows power-law, since there exists a dominant attractor ruin ($p_2 = 0.49$ and $q_2 = 0.99$).
However, the residence time distribution in the case of line (b) does not follow power-law.
In this case, one attractor ruin rarely attracts orbits, but is stayed by the orbits for a long time ($p_1 = 0.49$ and $q_1 = 0.01$).
Besides, another attractor ruin attracts frequently, but is stayed by the orbits for a short time ($p_2 = 0.25$ and $q_2 = 0.99$).
In this case the residence time distribution is multiscale.

On the other hand, if $p_i > 0.5$, then the probability $Q_i = \lim_{t \rightarrow \infty}Q(i,t)$ that transition at attractor ruin $i$ never happens is a positive value.
Since transition probability matrix $A$ is irreducible, the probability $R$ that orbits itinerate over attractor ruins forever is given by 
\begin{equation}
R = \lim_{n \rightarrow \infty} \big[ \sum_{i=1}^{M} q_i (1 - Q_i) \big]^n = 0.
\end{equation}
Thus, CI occurs only as a transient state in this case.

In the present paper, we have proposed the prototype model based on the CI mechanism.
By analyzing the model, we have shown that the residence time distribution averaged by all attractor ruins is the superposition of (truncated) power-law distributions.

\begin{figure}[hbtp]
\hspace{-100pt}(a) \newline
\vspace{-10pt}
\scalebox{0.6}{\includegraphics{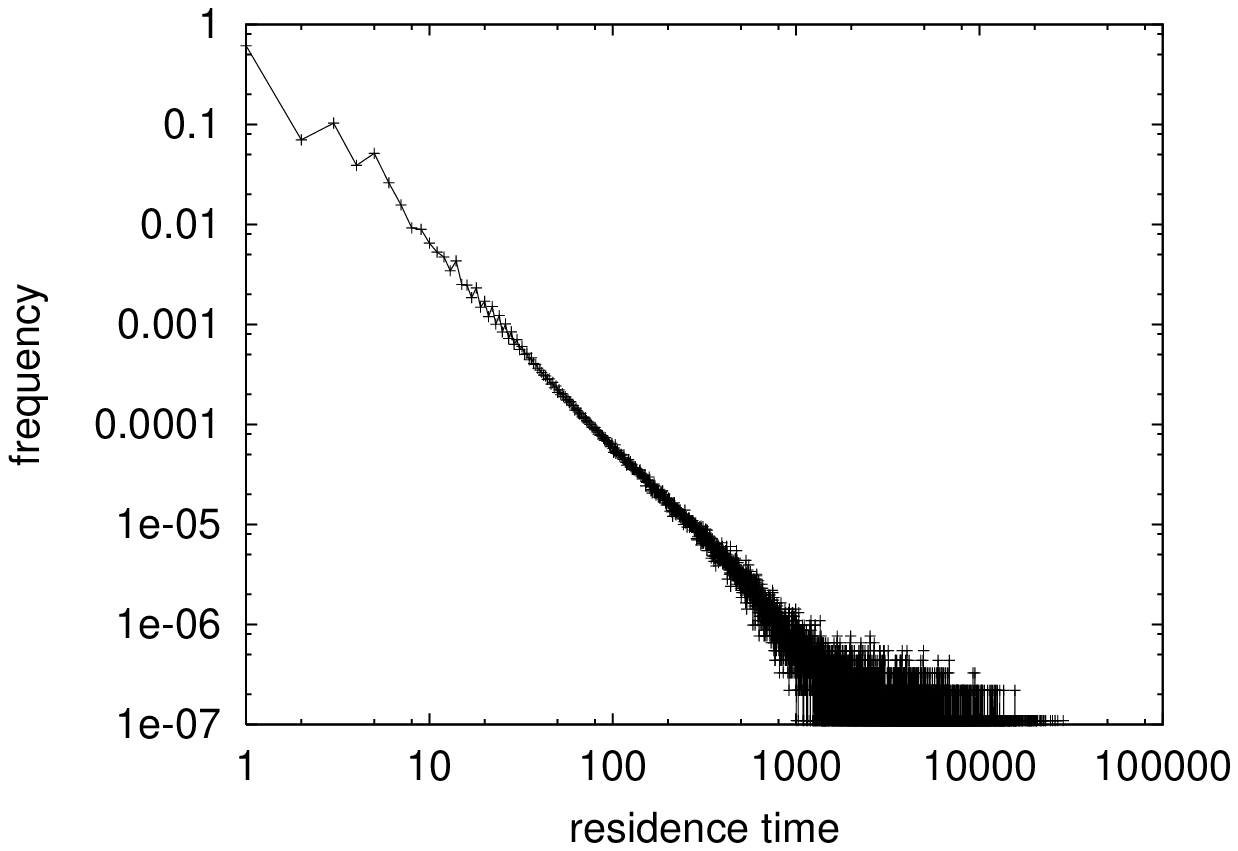}}

\vspace{10pt}
\hspace{-100pt}(b) \newline
\vspace{-10pt}
\scalebox{0.6}{\includegraphics{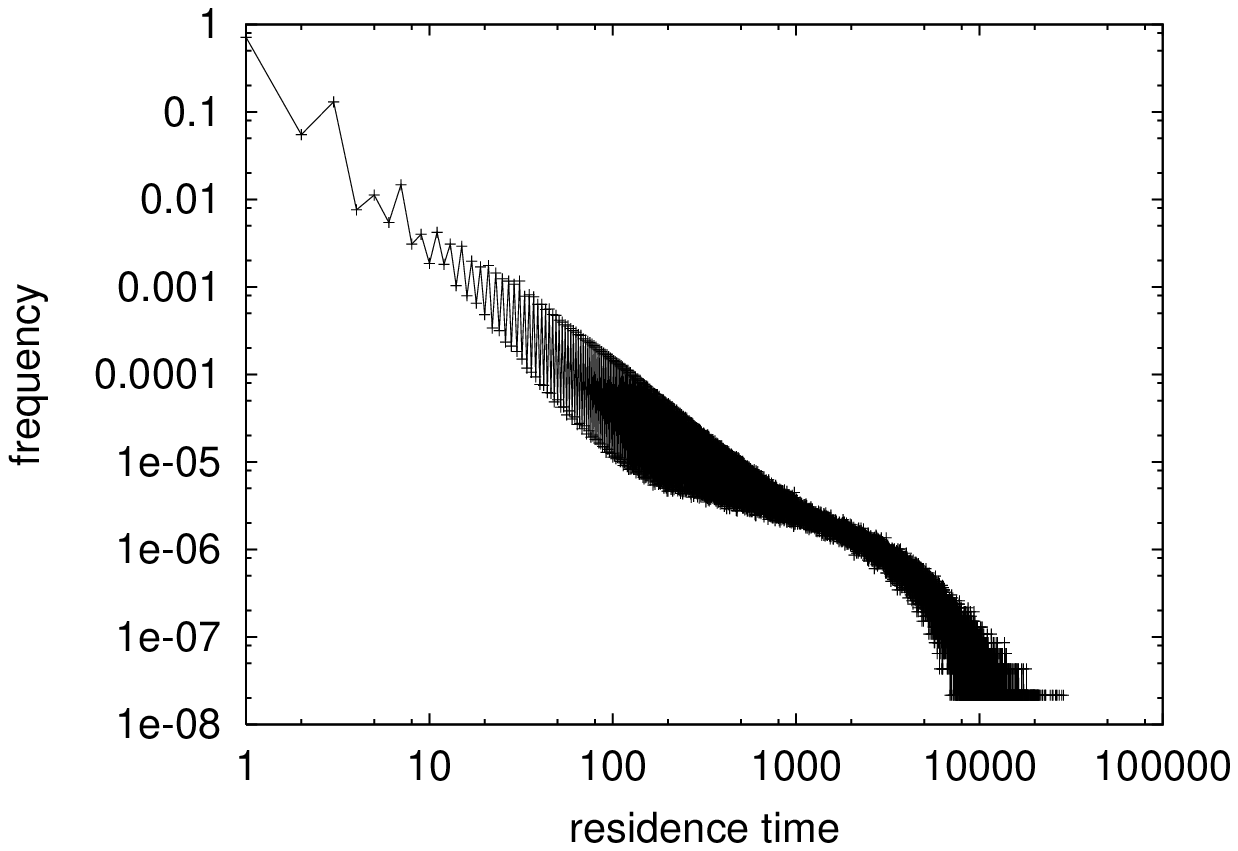}}
\caption{\small The residence time distribution averaged by all clustering conditions in GCM for $N = 10$.
In this simulation, elements $i$ and $j$ are synchronized if $|x_t(i) - x_t(j)| < 10^{-6}$.
(a)$\alpha = 1.57$ and $\epsilon = 0.3$;
(b)$\alpha = 1.9$ and $\epsilon = 0.2$.
}
\label{figure:global_residence_time_distribution_in_GCM}
\end{figure}

A computer simulation for GCM defined by Eq.(\ref{equation:globally_coupled_map}) was carried out to make sure of the result.
Figure \ref{figure:global_residence_time_distribution_in_GCM} displays the residence time distribution averaged by all clustering conditions in GCM with $f(x) = 1 - \alpha x^2$.
This clearly shows that the residence time distribution in GCM follows power-law as well as that in our model.
Furthermore, Fig.\ref{figure:global_residence_time_distribution_in_GCM}(a) implies that orbits itinerate over attractor ruins having almost the same distribution, and Fig.\ref{figure:global_residence_time_distribution_in_GCM}(b) implies with different distributions.

\begin{figure}[hbtp]
\hspace{-100pt}(a) \newline
\vspace{-10pt}
\scalebox{0.6}{\includegraphics{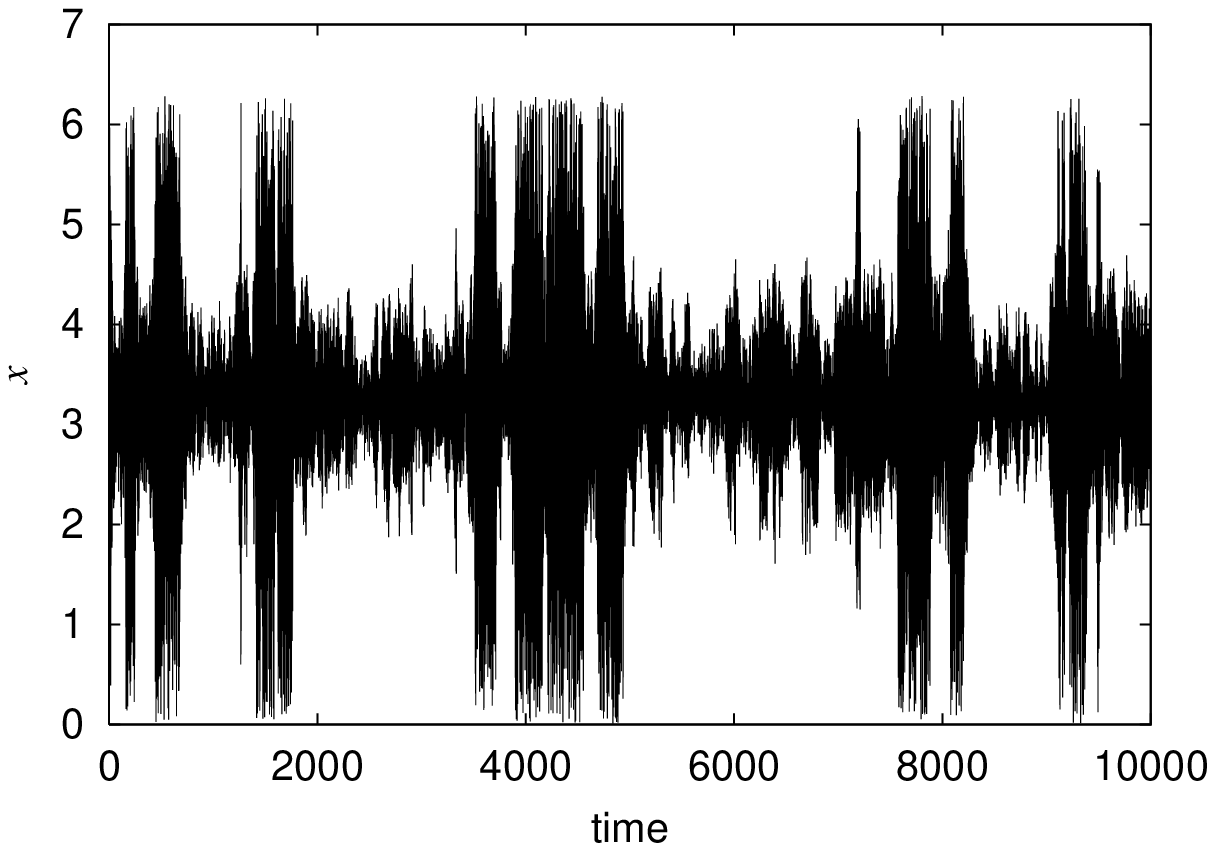}}

\vspace{10pt}
\hspace{-100pt}(b) \newline
\vspace{-10pt}
\scalebox{0.6}{\includegraphics{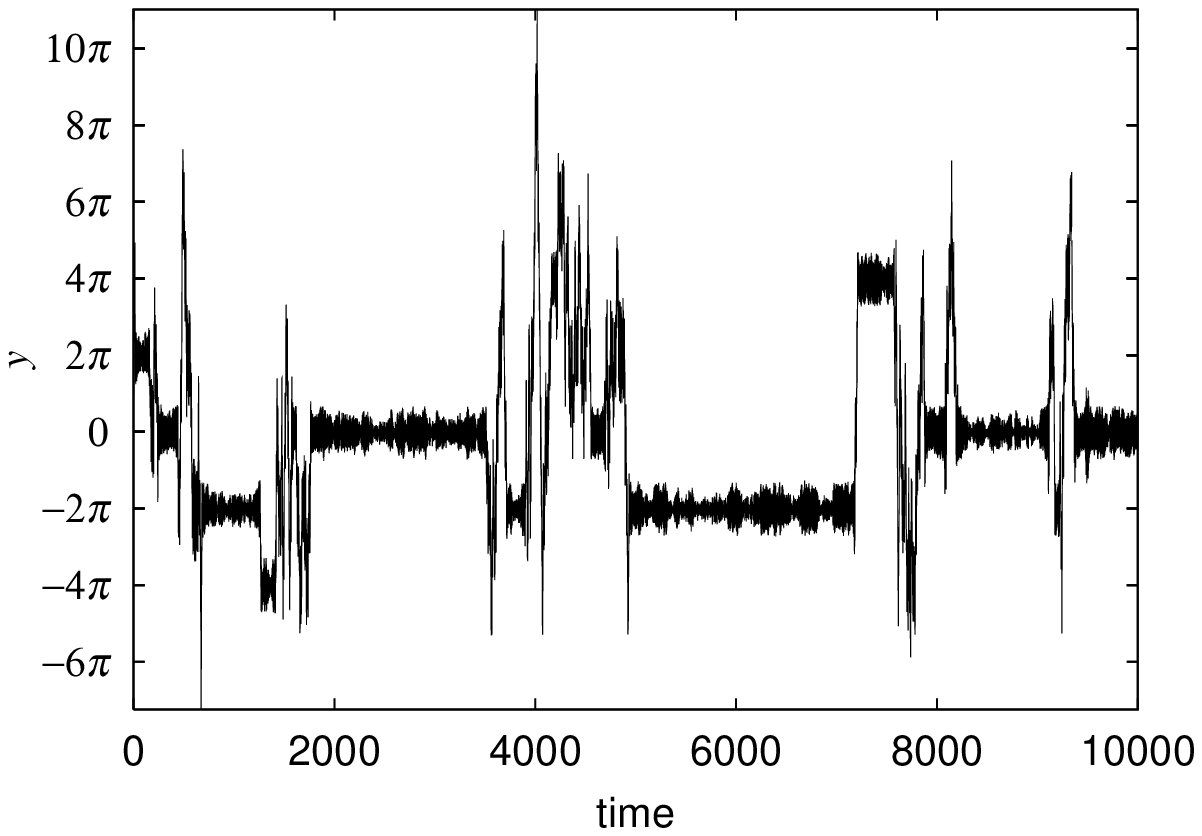}}
\caption{\small An example of the time sequence described by Eq.(\ref{equation:grebogi_model}) with $\nu = 0.02$, $\omega=3.5$, and $\sigma = 0.15$.
(a)phase of the rotor $x$;
(b)angular velocity $y$.
}
\label{figure:dynamics_of_Grebogi}
\end{figure}

\begin{figure}[hbtp]
\scalebox{0.6}{\includegraphics{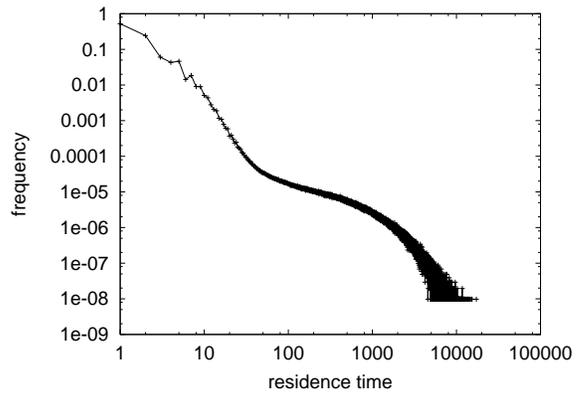}}
\caption{ The residence time distribution averaged by all attractor ruins given by Eq.(\ref{equation:grebogi_model}) with $\nu = 0.02$, $\omega=3.5$, and $\sigma = 0.15$.
Here each region corresponding to an attractor ruin is given by $(2n-1)\pi < y \leq (2n+1)\pi$ for each $n \in \mathbb{Z}$.
}
\label{figure:global_residence_time_distribution_in_Grebogi}
\end{figure}

As an another example of CI, we show dynamics of the kicked single rotor under the influence of noise\cite{KFG1999}.
This dynamics is defined by the following two-dimensional map:
\begin{eqnarray} \label{equation:grebogi_model}
x_{n+1} & = & x_n + y_n + \delta_{x} \quad \pmod{2\pi}, \nonumber \\
y_{n+1} & = & (1 - \nu)y_{n} + \omega sin(x_n + y_n) + \delta_{y},
\end{eqnarray}
where $x$ corresponds to the phase, $y$ corresponds to the angular velocity, the parameter $\nu$ is the damping, and $\omega$ is the strength of the forcing.
The terms $\delta_x$ and $\delta_y$, where $\sqrt{\delta_x^2 + \delta_y^2} \leq \delta$, are the amplitude of the uniformly and independently distributed noise.
The dynamics of Eq.(\ref{equation:grebogi_model}) is illustrated in Fig.\ref{figure:dynamics_of_Grebogi}.
As seen in Fig.\ref{figure:dynamics_of_Grebogi}(b), the orbit is attracted to certain ordered motion states for a while, and is kicked out of the states and behaves chaotically.
Figure \ref{figure:global_residence_time_distribution_in_Grebogi} shows a result of a computer simulation of Eq.(\ref{equation:grebogi_model}) for the residence time distribution of orbits at attractor ruins.
In this figure, the residence time distribution averaged by all attractor ruins seems to be the superposition of truncated power-law distributions as well as the line (b) in Fig.\ref{figure:global_residence_time_distribution}.
This figure and our results imply that two attractor ruins with different residence time distributions exist in the dynamics of Eq.(\ref{equation:grebogi_model}).
As the above experiment results show, theoretical results based on our model could be applied to other models showing CI.

Let us consider our result by comparing the mechanism of temporal intermittency in low-dimensional dynamical systems with that of CI.
Temporal intermittency is a phenomenon that bursts sometimes appear in the intervals of ordered states, and is seen at some points in the parameter space in the neighborhood of bifurcation boundaries\cite{BPV1984}.
Traditionally, researches of temporal intermittency discuss the occurrence of bursts and their intervals, while CI researches discuss chaotic itinerant motions among several ordered states.
Phenomenologically speaking temporal intermittency and CI share several features.
For example, ordered states and chaotic states appear by turns.
However, as discussed above, the mechanism of CI is different from that of classical temporal intermittency.
This difference appears as a difference in the residence time distributions\footnote{
Temporal intermittency can be classified into so called type-I, II, and III.
In type-I, maximum length of interval of bursts is given by $\epsilon^{-1/2}$, where $\epsilon$ is a positive value.
In type-II and type-III, probability distribution $P(t)$ of interval of bursts is either $P(T) \sim e^{-2\epsilon T}$ if $T \gg \epsilon^{-1}$ or $P(T) \sim (4\epsilon T)^{-3/2}$ if $1 \ll T \ll \epsilon^{-1}$.}.
Thus, it is necessary to consider CI as distinct from those temporal intermittencies.

In our model, if there is an attractor ruin $i$ with $p_i > 0.5$, it has been shown that CI occurs only as a transient state.
Note that $p_i > 0.5$ implies that an attractor ruin $i$ is a Milnor attractor, because a basin of attraction for the attractor ruin $i$ has nonzero measure.
Hence if there exists a Milnor attractor in our model, CI occurs only as a transient state.
On the other hand, the existence of CI in a coupled Milnor attractor system has been reported\cite{TU2003}.
From the present result, we can consider the following two possibilities for CI in such systems:
(a)the behavior like CI is strictly observed as a transient, or
(b)the transition probability among attractor ruins cannot be represented as a Markov chain.
If a system has a transition probability that cannot be represented as a Markov chain, it is expected that the system is more complex than our model.
Further analysis of such systems is required.

Finally, to improve our model in future, we address ideas for extending it.
In our model, the term of nonlinearity has been simplified as either $1$ or $-1$.
This restriction is easily removed, but we believe that we will obtain the same results even if we study such a general case.
As a more essential restriction, our model does not describe the concrete behavior of orbits on each attractor ruin.
While this simplification allows us to investigate easily the relationship between the stability and the transition probability of attractor ruins, we cannot discuss the dynamical behavior in each attractor ruin.
One of the extensions to express concrete behavior is that for any attractor ruins we prepare a function governing change of states.
If dynamics on an attractor ruin is determined by a function associated with such attractor ruin, we can represent the concrete behavior on each attractor ruin.
However, because orbits successively itinerate over attractor ruins, we must consider that the functions associated with attractor ruins dynamically change.
Functional shifts provide a framework to describe such dynamical systems\cite{NH2004}.
A functional shift is defined as a shift space that is a set of bi-infinite sequences of some functions on a set of symbols.
By using functional shifts, we can represent dynamical systems with dynamic change of functions.
Improving the model proposed in this Letter by using functional shifts will be the topic of a future study.

I am grateful to T. Hashimoto for useful discussions.
I would like to thank J. Steeh for critical reading of the manuscript.

\end{document}